\pdfoutput=1
\documentclass{article} % For LaTeX2e
\usepackage{nips14submit_e,times}
\usepackage{hyperref}
\usepackage{url}
\usepackage{graphicx}
\nipsfinalcopy

\title{A Model Selection Framework for Graph-based Data \thanks{This work was sponsored
by the Defense Advanced Research Projects Agency under Air Force
Contract FA8721-05-C-0002. Opinions, interpretations, conclusions,
and recommendations are those of the authors and are not necessarily
endorsed by the United States Government.}}

\author{
Rajmonda S. Caceres \\
MIT Lincoln Laboratory \\ \And
Leah Weiner \\ 
Brown University\\ \And
Matthew C. Schmidt \\ 
Laboratory for Analytical Sciences, NC State University \\ \And
Benjamin A. Miller \\ 
MIT Lincoln Laboratory\\ \And
William M. Campbell \\
MIT Lincoln Laboratory\\
}

\newcommand {\erdosrenyi}{Erd\"{o}s-R\'{e}nyi}

\begin{document}

\maketitle

\begin{abstract}
Graphs are powerful abstractions for capturing complex relationships in diverse application settings.  An active area of research focuses on theoretical models that define the generative mechanism of a graph. Yet given the complexity and inherent noise in real datasets, it is still very challenging to identify the best model for a given observed graph. We discuss a framework for graph model selection that leverages a long list of graph topological properties and a random forest classifier to learn and classify different graph instances. We fully characterize the discriminative power of our approach as we sweep through the parameter space of two generative models, the \erdosrenyi~and the stochastic block model. We show that our approach gets very close to known theoretical bounds and we provide insight on which topological features play a critical discriminating role. 
\end{abstract}

\section{Introduction}
\label{intro}
Across many application areas such as social media, biology and cyber, we observe the generation of large amounts of data, with millions or billions of entities (e.g., people, proteins or IP addresses) and many complex relationships among them. This continuous data growth has sparked an ongoing interest in the ability to model and extract useful information from such data. Graphs are often a natural abstraction, where each entity in the data is represented by a node and a relationship between entities is represented by an edge. There has been a lot of research towards modeling and estimating the underlying graph generative mechanisms. The mapping of an observed graph instance to a model allows us to apply the theoretical knowledge we have about the model and to make precise claims about the underlying structure  of the data. However, existing generative models make many simplifying assumptions and often it is not clear how these assumptions affect the representative power of the model. 

As an alternative, we consider the problem setting where rather than assuming one graph model for the data and estimating its parameters, we would like to select the best model given a set of candidates. In this paper, we discuss a framework for mapping unlabeled graph instances to the closest generative model leveraging  a comprehensive list of topological features.  We focus our analysis on sparse graphs, because real world graphs tend to be very sparse, and because in many problem settings sparsity tends to complicate various learning tasks. We concentrate on the following questions: 1) Can we discriminate graph models in the sparse regime? 2) How does discrimination power deteriorate as a function of structure and noise? 3) What are the critical features for discrimination? 

Previous literature has looked at the model selection problem for graph data. A variety of feature types have been considered to represent the graph, including topological features~\cite{Airoldi2011}, distributions of frequent subgraphs~\cite{Janssen2012,Middendorf2004}, and spectral features of graph matrices~\cite{Fay2011,Takahashi2012,Zhu2005}. These related studies are primarily focused on small, dense graph instances and only consider a few parameter choices for several graph models such as the \erdosrenyi, scale free and small-world models.

Our study differs from previous research in four important ways: 1) We are interested in the characterization of model selection power for the full parameter space, not just specific parameter choices 2) We focus on analyzing sparse graphs (graphs with $\mathcal{O}(V)$ number of edges), since sparsity is a typical feature of observed real world graphs, but has not been thoroughly analyzed in previous literature (3) We analyze the Stochastic Block model, a model that similarly has not received a lot of attention in the past, yet is very useful in explicitly defining community structure (4) We use random forest (RF)~\cite{Breiman2001}, an ensemble classifier known for its robustness when handling noisy, highly dimensional datasets. As we demonstrate in Section~\ref{results}, an RF classifier trained on topological features achieves close to optimal classification performance and maintains its robustness under various noise settings.

\section{Methodology}

\begin{figure}[h]
\begin{center}
\includegraphics[width=.8\columnwidth]{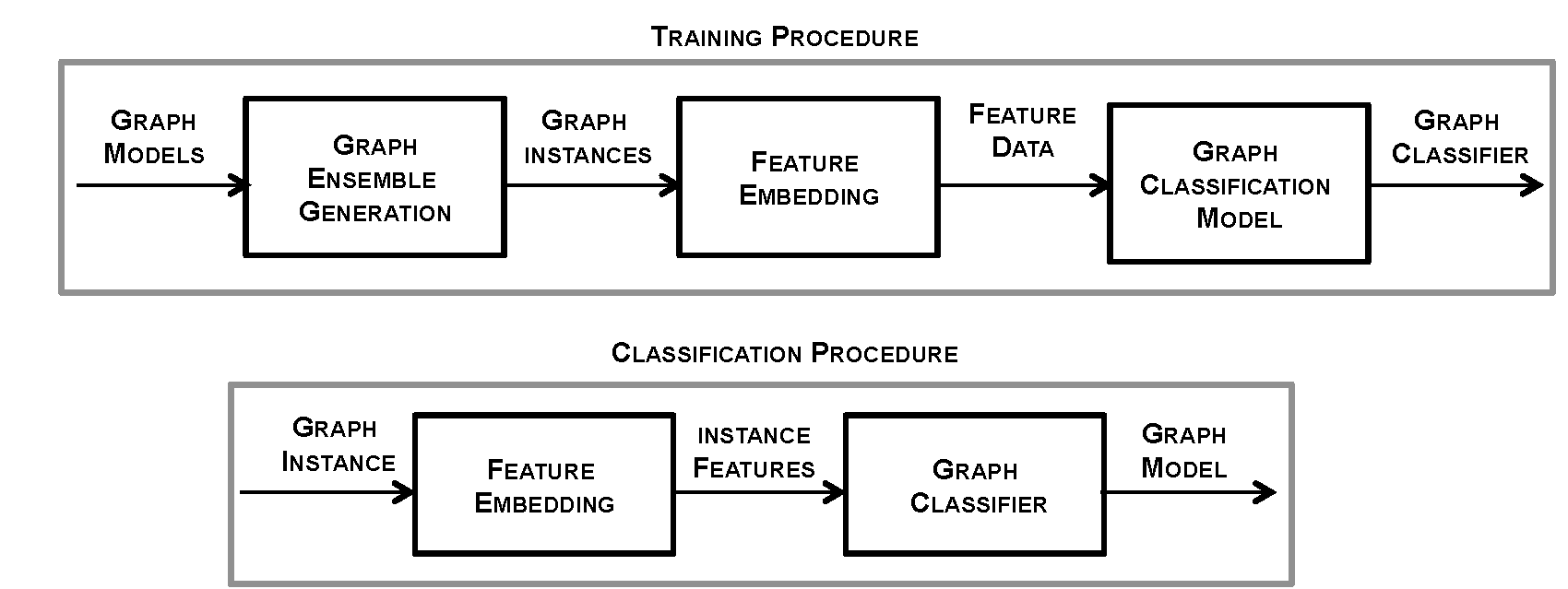}
\end{center}
\caption{Graph Model Selection Framework.} 
\label{fig:framework} 
\end{figure}

We extend the graph model selection framework previously used in~\cite{Airoldi2011,Janssen2012,Middendorf2004}. A diagram of this framework is shown in Figure~\ref{fig:framework}. We consider the discrimination of two graph models: the \erdosrenyi~\cite{Erdos60} and the stochastic block model~\cite{Snijder1997}. The \erdosrenyi~model describes a scenario where each pair of entities connects with equal probability $p$. The stochastic block model describes a more realistic scenario where entities are organized into non-overlapping communities (blocks) and the probability of entities connecting depends only on their community affiliation. Entities within the same community connect with probability $p_{in}$, while entities in different communities connect with probability $p_{out}$, where $p_{in} >p_{out}$.  For each these models and for density values in the sparse regime, we generate an ensemble of graph instances. A detailed description of the parameters considered in our experiments is summarized in Table~\ref{table:param}. 

We then embed each graph instance into a topological feature space. We consider the following features\footnote{Note that if one feature is assigned four numbers, this means we considered the maximum, the minimum, the average and the standard deviation over each vertex in the graph instance}: degree centrality (1-4), betweeness centrality (5-8), closeness centrality (8-12), clustering coefficient (13-16), diameter (17), radius (18), triad count (19-22), average shortest path length (22-25). Finally we train an RF classifier and use the learned model to classify a collection of unlabeled graph instances. 

\section{Experimental Results}
\label{results}
\subsection{Characterization of discrimination power across the parameter space}
We are interested in characterizing the ability to discriminate an \erdosrenyi~graph from a stochastic block graph as a function of their parameters. We quickly found that when the (expected) density of the \erdosrenyi~graph was not equal to the density of the stochastic block graph, discrimination was trivial. Therefore, we focused our efforts on distinguishing graph instances of equal density. 

\begin{figure}[h]
\begin{center}
\includegraphics[width=.8\columnwidth]{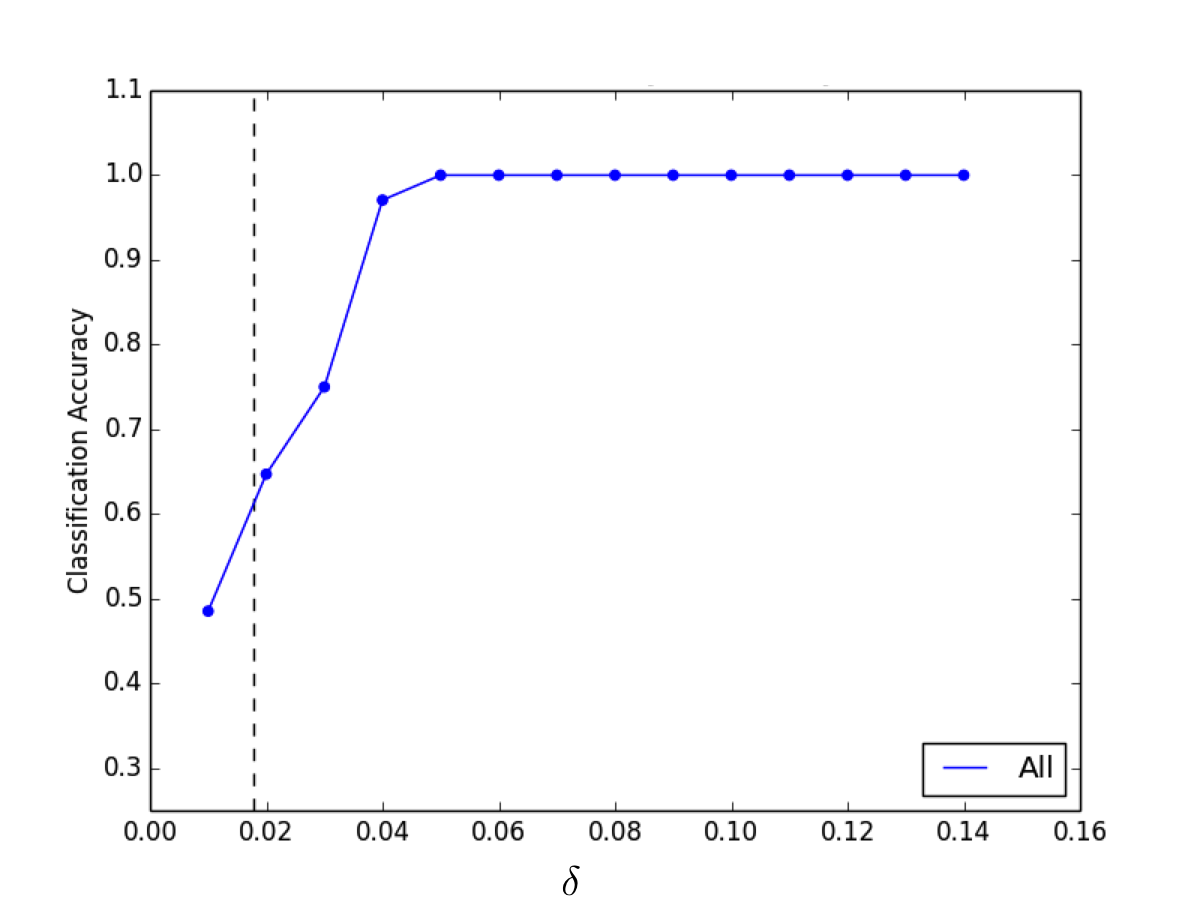}
\end{center}
\caption{RF performance as a function of $\delta=p_{in}-p_{out}$ for  \erdosrenyi~and stochastic block graph instances with density 0.08. The dashed line represents the theoretical value $\delta^*$ beyond which the stochastic block model becomes indistinguishable from the \erdosrenyi~model.} 
\label{fig:rf_acc} 
\end{figure}

The random forest classifier was trained on 66 instances of each model and tested on 34 instances of each model. Figure~\ref{fig:rf_acc} shows its performance for graphs of density $.08$ (i.e. $p=\frac{p_{in}+p_{out}}{2}=.08$). The x-axis represents the signal to noise gap $\delta=p_{in}-p_{out}$, which is a common way to measure community strength in theoretical settings. The y-axis measures the RF classification accuracy. The vertical dashed line represents the theoretical value $\delta^*= \sqrt{\frac{2(p_{in} + p_{out})}{n}}$ beyond which no algorithm is able to distinguish between the two graph models~\cite{nadakuditi2012}. 
As expected, when $\delta$ gets smaller and the community structure in the stochastic block model becomes less pronounced, the RF classifier performance drops accordingly. What is interesting is that the RF performance degrades in a similar way to what is characterized analytically in~\cite{nadakuditi2012}. We observe the same behavior across different graph density values.

\begin{table}[t]
\begin{center}
\begin{tabular}{ll}
\multicolumn{1}{l}{\bf Model Type}  &\multicolumn{1}{l}{\bf Parameters}
\\ \hline \\
\erdosrenyi         & $n=1000,p=.01,.015, \ldots,.09$\\
Stochastic block  & $n=1000,p_{in}=.01,.015, \ldots,.19$, $p_{out}=.01,.015, \ldots ,.09$ \\
\end{tabular}
\end{center}
\caption{Summary of model parameters considered in our experiments.}
\label{table:param}
\end{table}

\subsection{Identification of discriminatory features}
\begin{figure}[h]
\begin{center}
\includegraphics[width=.497\columnwidth]{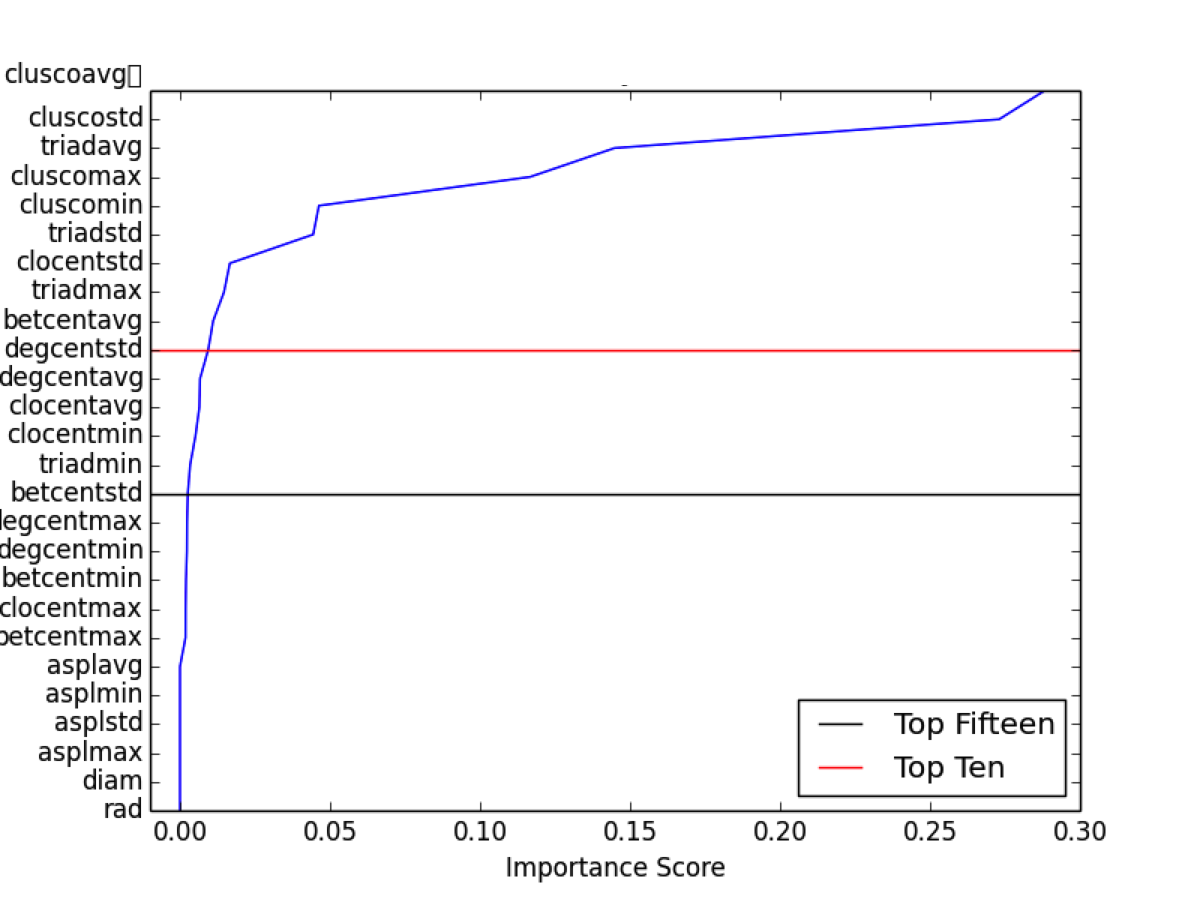}
\includegraphics[width=.497\columnwidth]{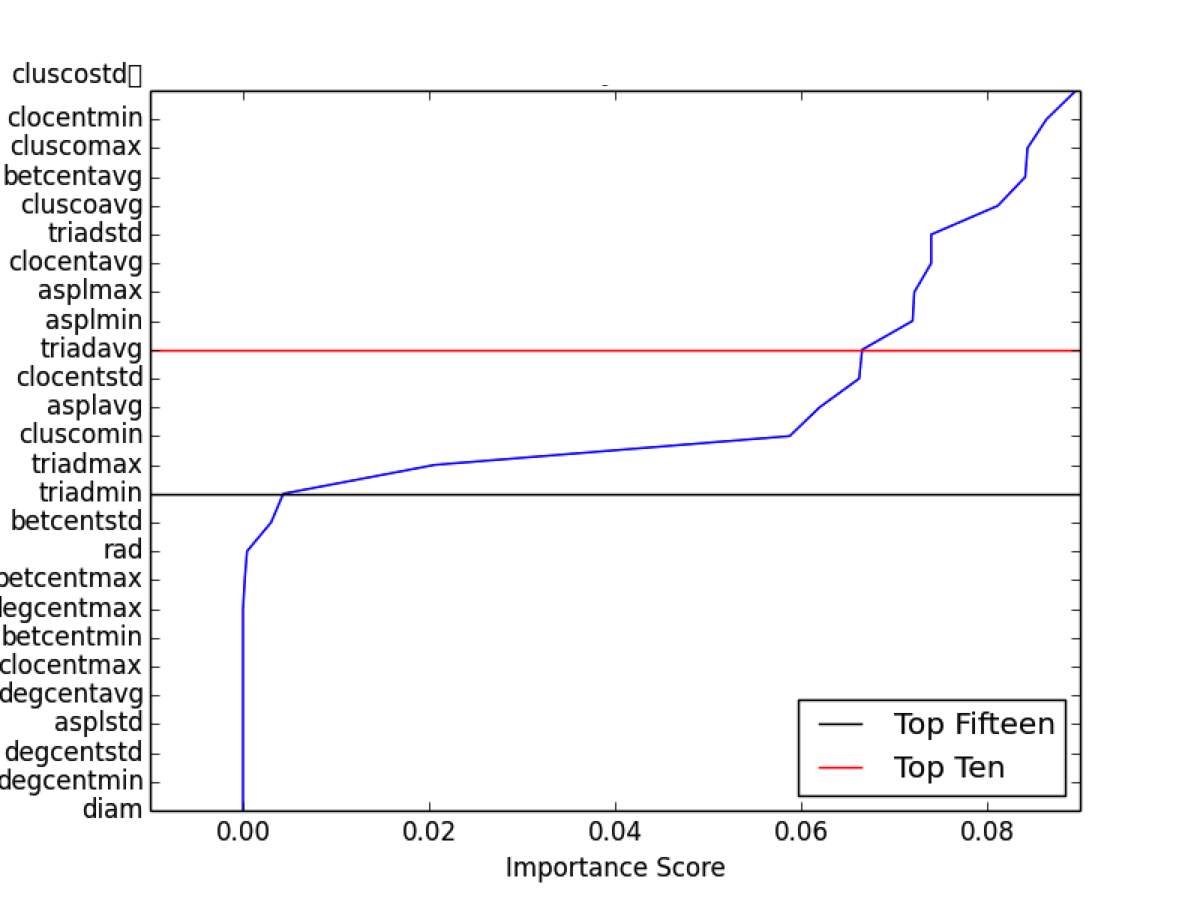}
\end{center}
\caption{Feature importance rankings for two levels of community structure (strong community structure on the left and weak community structure on the right) and fixed density of $.08$.} 
\label{fig:featImp} 
\end{figure}

The random forest method ranks the input features based on how critical they are in discriminating the two models. Figure~\ref{fig:featImp} shows this ranking for various levels of problem difficulty (i.e: various levels of community strength in the stochastic block instances). We observe that regardless of the level of discrimination difficulty, about 15 out of 26 of the original features seem to capture most of the discriminative power of the classifier.

\begin{figure}[h]
\begin{center}
\includegraphics[scale=0.5]{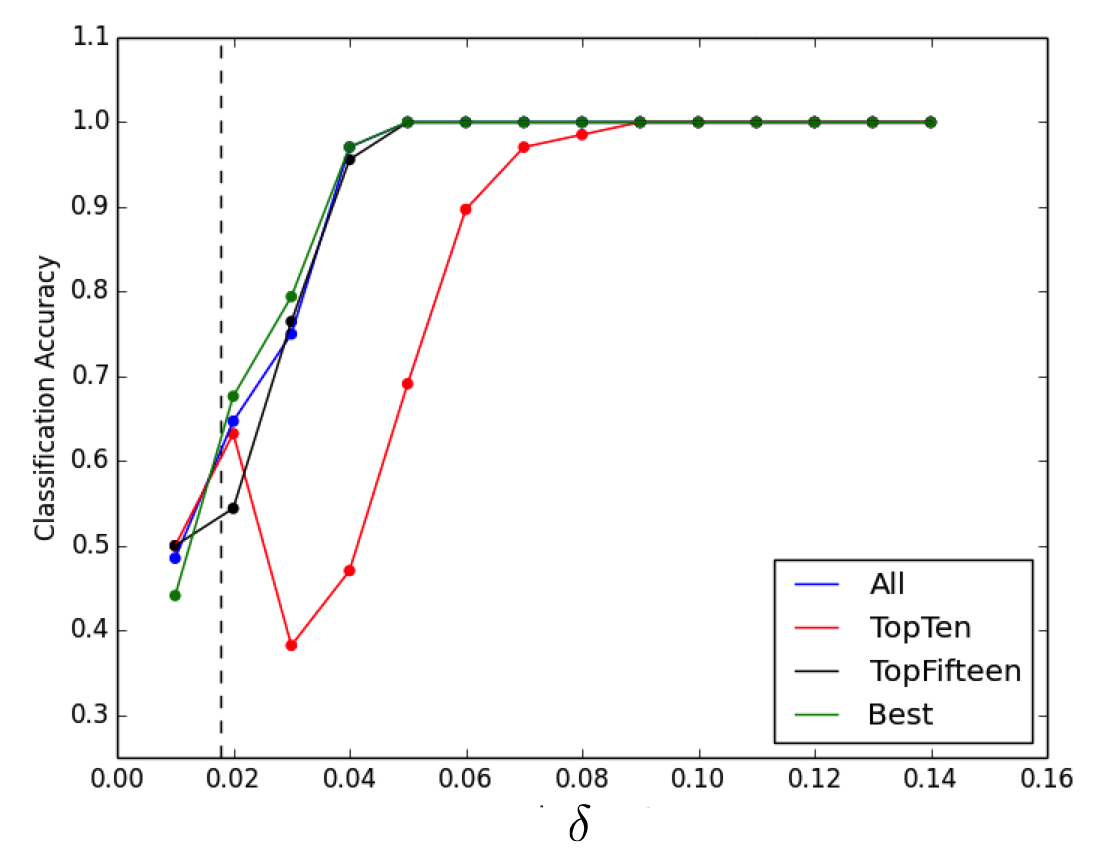}
\end{center}
\caption{RF classifying performance as a function of the gap $\delta=p_{in}-p_{out}$ for graph instances with density .08, using different feature subsets.} 
\label{fig:rf_accTopFeaturesBest} 
\end{figure}

Based on this observation, we retrain the RF classifier using only the top 10 and top 15 performing features. The classification accuracy for these experiments is shown in Figure~\ref{fig:rf_accTopFeaturesBest}. We can see that using only the top fifteen features results in a comparable level of accuracy as using all 26 original features. However, using only ten features seems to greatly affect the discrimination power of the classifier. These results are illustrated for graphs of density .08, but hold across all density values we considered.

It is important to note that the set of top fifteen features is not the same for different parameter choices. This seems to indicate that for different levels of problem difficulty, different topological features become critical for model selection. There are however some patterns that arise. Among the features that tend to be important in many parameter settings are betweenness centrality (min, max), closeness centrality (min, max, avg, std), clustering coefficient (min, max avg, std), triad count (max, avg, std) and average shortest path (max, avg, std). As shown in Figure~\ref{fig:rf_accTopFeaturesBest}, when re-training the RF classifier just on these features (green line on the plot), we obtain nearly identical classification results. 

\subsection{Robustness of random forest classifier}
\begin{figure}[h]
\begin{center}
\includegraphics[width=.497\columnwidth]{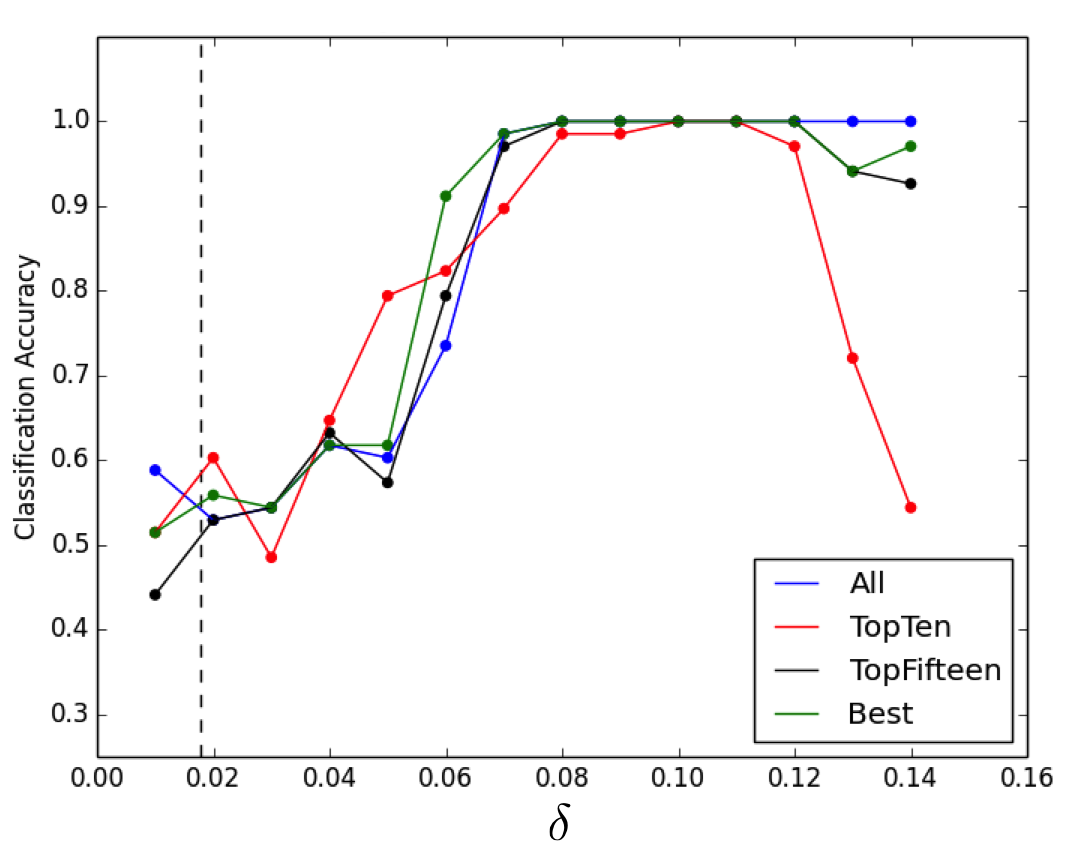}
\includegraphics[width=.497\columnwidth]{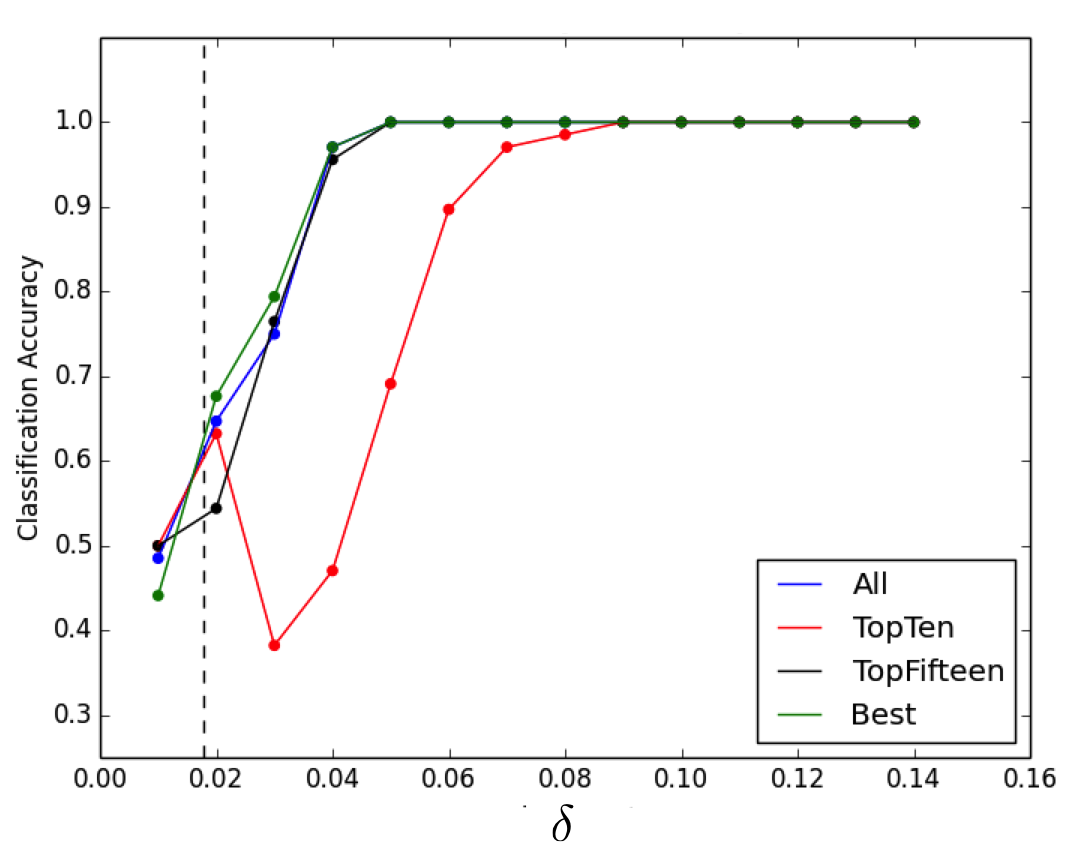}
\end{center}
\caption{RF accuracy for density $0.08$ when uniformly rewiring 10\% (left) and 20\% (right) of the edges.} 
\label{fig:rf_rewire} 
\end{figure}

So far we have analyzed classification performance based on pure graph instances, that is, graph instances generated following one of the generative models. In realistic settings, we expect graph instances to contain noise. To this end, we generated noisy graph instances by performing random edge rewiring at different levels. Figure~\ref{fig:rf_rewire} shows the RF classifier performance when tested on graph instances with 10\% and 20\% of the edges re-wired. Note that uniformly rewiring edges of stochastic block graph instances results in graphs with less pronounced community structure. The rewiring, therefore, has the effect of making a stochastic block graph instance more \erdosrenyi-like and therefore it increases the discrimination difficulty. We observe that when we rewire only 10\% of the edges, while the level of discrimination decreases, the same performance trend is present and we can still discriminate relatively well. However, when we rewire 20\% of the edges, discrimination power drops significantly. One interesting thing we observe is that using only the top ten features makes our classifier more robust. This result contrasts what we observed from experiments on pure graph instances, where using only the top ten features did not result in good discrimination power.

\begin{figure}[h]
\begin{center}
\includegraphics[width=.497\columnwidth]{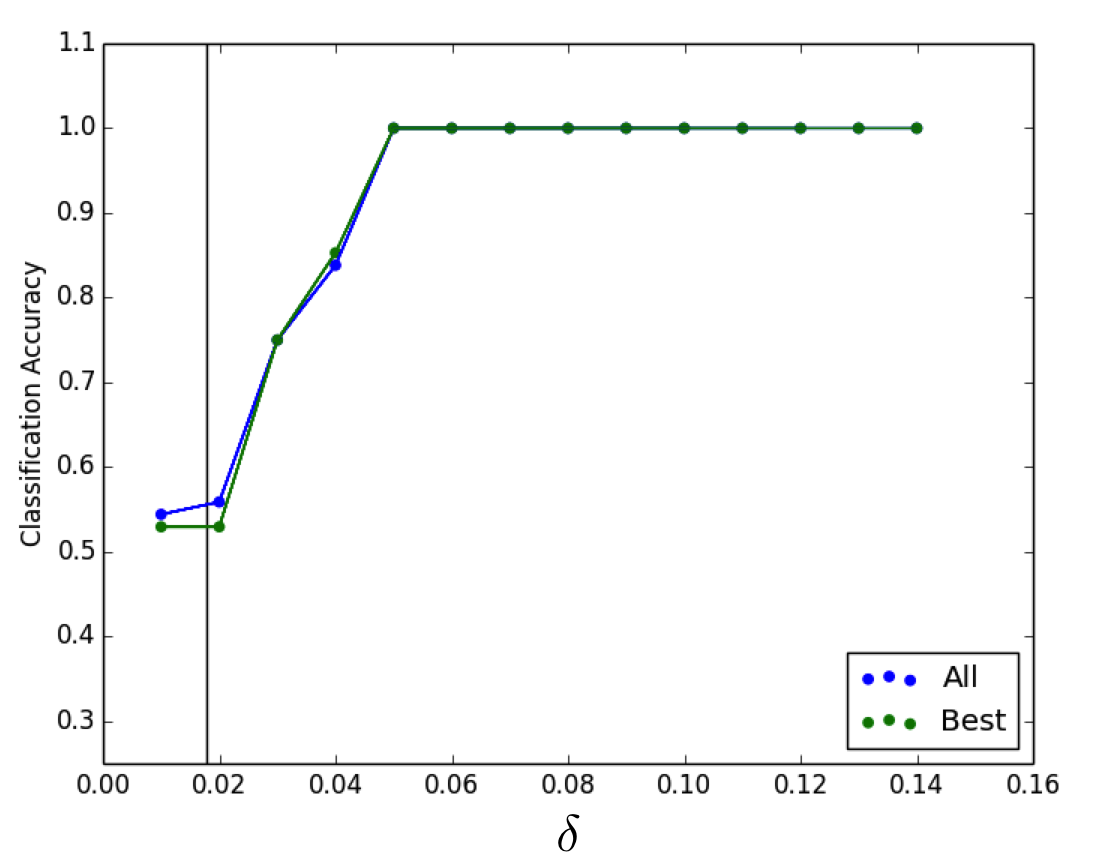}
\includegraphics[width=.497\columnwidth]{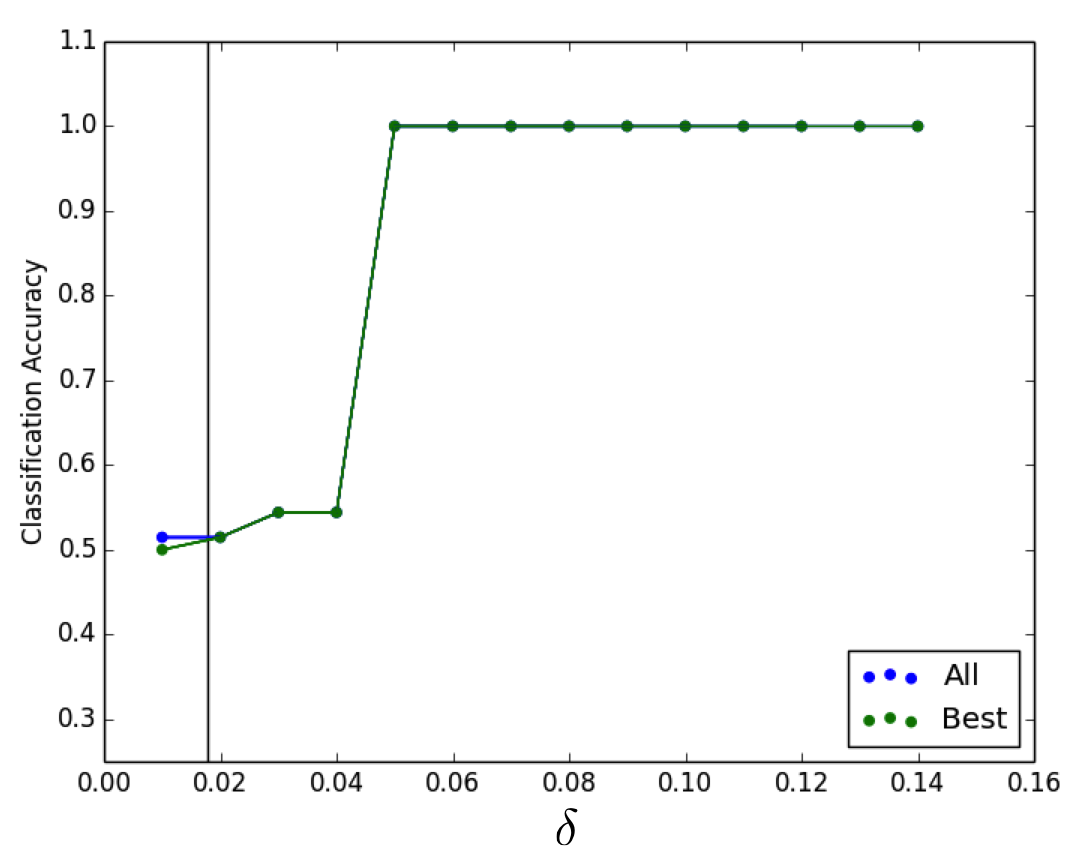}
\end{center}
\caption{RF Classification accuracy when aggregating graph instances with parameter errors at most .02 (left) and at most .04 (right)  for graph instances of density .08.} 
\label{fig:agg} 
\end{figure}

In all the experiments we have performed thus far, we assume the model parameters for the unlabeled graph instances are known and our only task is to pick one model over the other. In realistic scenarios however, we can cannot make this assumption and an additional parameter estimation step is required before we perform model selection. Parameter estimation is often an imperfect process, either due to the noise in the graph instance itself, or the errors of the estimation method. For this reason, we were interested in understanding how much parameter estimation error can the RF classifier handle while still maintaining a good classification performance. To answer this question, we 
performed the following experiment: for the stochastic block model we created training ensembles by considering graphs instances of the same density but not necessarily the same parameters $p_{in}, p_{out}$. That is, as long as the density is the same, stochastic block model instances of different parameters were labeled the same during the training process. During testing however we labeled the graph instances using their unique parameter values. This experiment is designed to identify regions in the parameter space that lead to similar discrimination performance. Figure~\ref{fig:agg} illustrates the RF classification results of this experiment. We observe that the RF classifier is tolerant to parameter errors of $.02$, but starts deteriorating considerably for parameter values bigger than .04.

\subsection{Graph size consistency properties}
\begin{figure}[h]
\begin{center}
\includegraphics[scale=0.5]{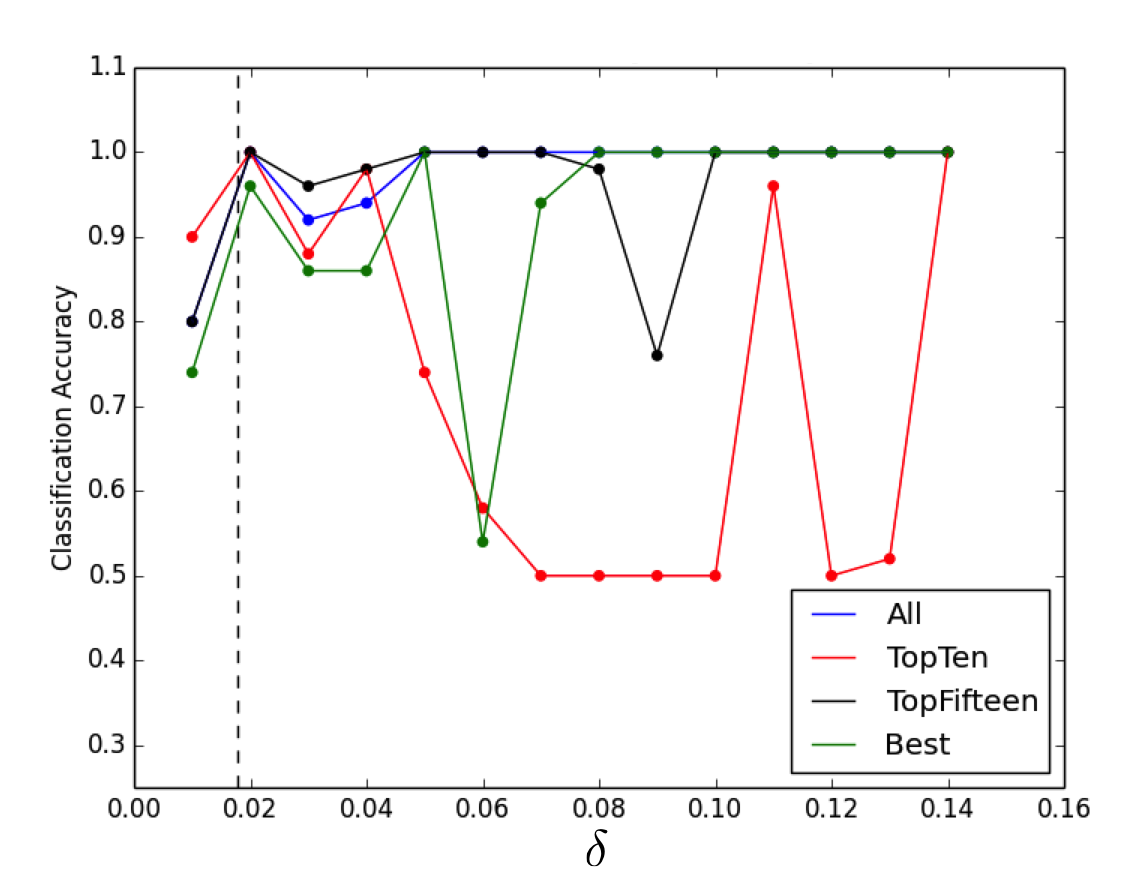}
\end{center}
\caption{Classification accuracy when RF classifiers are trained on graphs of size n=1000, but tested against graphs of size $n=1100$.} 
\label{fig:n1100} 
\end{figure}

In many realistic settings, we observe or analyze graphs of different sizes that are generated by the same underlying mechanism. In such scenarios, we would like a model selection method that is not sensitive to the size of the graph. To analyze sensitivity of out approach to the size of the graph, we tested RF classifiers trained on $1000$ vertex graphs, against graph instances of larger size, but same parameters $p,p_{in},p_{out}$. Figure~\ref{fig:n1100} shows RF classification results when we test on graphs with n=1100. We observe that RF classification accuracy degrades for sparser graphs, but remains relatively robust (and even improves in some cases) in the denser parameter space. Similar to previous experiments, we observe that about 15 of the original features carry most of the discriminative power. However, dropping down to the top 10 features severely reduces classification accuracy.

\section{Conclusions}
Our experimental study shows that a random forest classifier is able to almost perfectly separate a sparse stochastic block model graph from a sparse \erdosrenyi~graph when compared against known theoretical performance bounds. The RF classifier maintains good classification performance under various noise settings such as erroneous parameter estimation and random edge rewiring. A complicated picture emerges when analyzing the topological features that play a critical role in discriminating the two models. Even though, a smaller set of about 15 of the original 26 features are sufficient to maintain the same discrimination accuracy, this number is still quite large considering the small number of parameters needed to generate the two models. Furthermore, at different points in the parameter space, different topological features emerge as critical for model selection. We also observe the same behavior when the RF classifier is tested under noisy samples. This complex behavior seems to reinforce the intuition that graph objects are nontrivial to embed and that special care is needed when choosing graph feature representations. In the future, we plan to further analyze the RF classifier sensitivity to a range of graph sizes as well as apply the same model selection framework to additional graph models.

\bibliographystyle{plain}
\bibliography{related}
\end{document}